To quote this work :

V. Brien, A. Dauscher, F. Machizaud
Optical reflectivity as a simple diagnostic method for testing structural quality of icosahedral quasicrystals
Journal of Applied Physics A , 100(4): 43503-1-8 (2006)
hal-02088173, doi/10.1063/1.2234558

Thank you

---

# Optical reflectivity as a simple diagnostic method for testing structural quality of icosahedral quasicrystals

Valérie Brien, Dr. (Corresponding author)

Valerie.Brien@lpmi.uhp-nancy.fr, Tel : +33 (0)3 83 68 49 28, Fax : +33 (0)3 83 68 49 33

*Laboratoire de Science et Génie des Matériaux et de Métallurgie, UMR CNRS-INPL 7584, Parc de Saurupt, ENSMN, F-54042 Nancy Cedex, France*

*Now at : Laboratoire de Physique des Milieux Ionisés et Applications, UMR CNRS-UHP 7040, Université Henri Poincaré Nancy, Faculté des Sciences et Techniques, B.P. 239, F-54506 Vandoeuvre-lès-Nancy Cedex, France,*

Anne Dauscher, Dr.

dauscher@mines.inpl-nancy.fr, Tel : +33 (0) 3.83.58.41.70

*Laboratoire de Physique des Matériaux, UMR CNRS-INPL-UHP 7556, Parc de Saurupt, ENSMN, F-54042 Nancy Cedex, France*

and

Francis Machizaud, Pr.

*Laboratoire de Science et Génie des Matériaux et de Métallurgie, UMR CNRS-INPL 7584,*

Parc de Saurupt, ENSMN, F-54042 Nancy Cedex, France, Now retired.

The optical reflectivity of Al-based and Ti-based quasicrystalline and approximant samples were investigated versus the quality of their structural morphology using optical reflectometry, X-ray diffraction and transmission electron microscopy. The different structural morphologies were obtained using three different preparation processes : sintering, pulsed laser deposition and reactive cathodic magnetron sputtering. The work demonstrates that the canonical behaviour of icosahedral state in specular reflectivity is extremely sensitive to different and very fine aspects of the microstructure : sizes of grains smaller than 50 nm, slight local diffuse disorder and shifts away from the icosahedral crystallographic structure (approximants). The work explains why the optical properties of the same kind of quasicrystals found in literature sometimes reveal a different behaviour from one author to another. The study then confirms the work of some authors and definitely shows that the canonical behaviour of icosahedral state in specular reflectivity over the 30000 - 50000 $cm^{-1}$ domain is characterized by a decreasing function made of steps. It also shows that this behaviour can be interpreted thanks to the cluster hierarchy of the model of Janot.

## I. INTRODUCTION

In 1984, the icosahedral crystallographic structure was discovered in the Al-Mn system.[1] Since then, many other phase diagrams showed to exhibit the same quasicrystalline structure (Al-Pd-Mn, Al-Cu-Fe, Al-Cu-Fe-B, Ti-Ni-Zr, etc …) Quasicrystalline structures exhibit an orientation-ordered structure with classically forbidden rotation symmetries. They possess an absence of periodic translation ordering. The new class of materials exhibits a new combination of properties. Indeed electronic, thermal, optical, tribological properties have been widely investigated in the past decades and were found to be very original with respect to what could have been expected regarding the physical and chemical behaviour of their metallic components. [2-12] Some crystals with big lattice parameters called approximants have a local order which is very close to the one of some quasicrystals. The structural similarity between quasicrystals and their approximants makes their physical and chemical properties close from each other. Within the mass of the work made on the physical characterization of these materials, optical properties were found to be rather difficult to be interpreted. For instance, Karpus *et al.*[13] underlined that the measured data were complex and stressed on the importance of the quality of the samples. This is of course well known from all material scientists and physicists : purity and structural quality of samples are of high relevance to identify their canonical physical properties. A right interpretation of optical properties of the icosahedral phases and their closely related phases cannot be done with no care for these considerations. Indeed, although some significant progress seems to have been done recently in the field, [14-16] the authors think it is pertinent to be more specific about the influence of the quality of quasicrystalline samples and their approximants on optical reflectivity, especially in the ultra-violet and far ultra-violet domains.

In this paper we report on the preparation and on the optical properties of several icosahedral (Al-Cu-Fe, Al-Cu-Fe-B, Al-Pd-Mn, Ti-Ni-Zr) and approximant (Al-Cr-Fe, Al-Cr-Fe-Mo) samples of different structural quality. The different qualities of the quasicrystalline samples were obtained either by preparing samples of the same system by two distinct synthesis processes, either by changing the preparation parameters using one synthesis process.

The structural quality of the samples was evaluated by X-ray diffraction (XRD), transmission electron microscopy (TEM) and scanning electron microscopy (SEM) when necessary.

Optical specular reflectivity of all samples was measured and is linked to the structural quality of the samples. The paper insists on the very high sensitivity of optical reflectivity data in the ultra-violet (UV) domain to the degree of structural order of the samples. The optical behaviour of several systems was considered. ~~Results of~~ The reflectivity response of the Al-Cu-Fe-B, Al-Pd-Mn system on all the frequency domain could be found in [15] , [15,17], respectively. Partial optical data of the Al-Cu-Fe system could be found in Homes *et al.*[18] and Eisenhammer *et al.*,[19] Ti-Ni-Zr optical data are new. Optical data of approximant AlCrFe-γ (γ brass phase) was already published.[15] The data of the $O_1$ phase (AlCrFe-$O_1$ and AlCrFeMo-$O_1$) are new although data obtained on a ($O_1$/$O_2$)Al-Cr-Fe mixed sample was presented in [15].

## II. SAMPLE PREPARATION

Bulk samples or thick films were prepared according to three different manufacturing processes : sintering of either powdered crushed ingots, or atomised powder, pulsed laser deposition *(PLD),* and reactive magnetron sputtering. For each process, several samples were prepared with technical process parameters close one from each other. They are compiled in tables 1 and 2.

Al-Cu-Fe-B and Al-Pd-Mn quasicrystalline samples and Al-Cr-Fe and Al-Cr-Fe-Mo approximant samples were prepared according to a classic sintering process. Three Al-Cu-Fe-B samples of the same composition $Al_{58.4}Cu_{25.1}Fe_{12.5}B_4$ (named AlCuFeB-0, AlCuFeB-1, AlCuFeB-2) were prepared. AlCuFeB-0 was prepared by hot-press ~~sing~~ sintering a quasicrystalline atomised powder whose particle size was around 25 μm, at 1103 K under 40 MPa uniaxial stress, followed by a post annealing treatment at 1053 K for 3 hours to eliminate the non-wanted CsCl-type β phases. AlCuFeB-1 and AlCuFeB-2 samples were prepared from powders whose particle size was taken in the 25-65 μm range, obtained by crushing ingots. The ingots were previously obtained by melting the pure constitutive elements by induction under a helium atmosphere in a copper mould crucible using r-f energy. The sintering treatment of AlCuFeB-1 (respectively AlCuFeB-2) was 1053 K under 15 MPa with a post annealing treatment of 3 hours at 373 K (respectively 903 K). Slow cooling to room temperature at 20 K/min followed, while maintaining the applied pressure. Two Al-Pd-Mn samples of $Al_{69.8}Pd_{21.6}Mn_{8.6}$ composition (named AlPdMn-0 and AlPdMn-1) were prepared. AlPdMn-0 and AlPdMn-1 samples were prepared from crushed ingots prepared as described above. ~~in a similar way as just~~. Thermal sintering treatment plateau of AlPdMn-0 (respectively AlPdMn-1) was set at 1103 K and followed by an annealing treatment for 3 hours at 873 K (respectively 373 K).

Three Al-Cr-Fe and Al-Cr-Fe-Mo samples were also prepared from crushed ingots : a sample named Approx-0 of $Al_9(Cr,Fe)_4$ composition, a sample named Approx-1 of $Al_{71.6}Cr_{22.4}Fe_6$ composition and a sample named Approx-2 of $Al_{71.6}Cr_{20.8}Fe_{5.9}Mo_{1.6}$ composition. Thermal sintering treatment plateau of 30 min was maintained at 20 K under the melting temperature of the approximant (i.e. 1333 K for the Approx-0 $\gamma$ brass phase and 1253 K for the Approx-1 and Approx-2 phases) followed by an annealing treatment for 2 hours at 373 K.

Two Ti-Ni-Zr thick films of $Ti_{41.5}Ni_{17}Zr_{41.5}$ composition (named TiNiZr-0 and TiNiZr-1) were prepared by pulsed laser ablation using a pulsed Nd : YAG (Qantel, YG 571C) laser. The wavelength used was 1064 nm, the repetition rate was 10 Hz and the pulse durations was 10 ns. The laser fluence used was 72 J/cm$^2$. Experiments were conducted under high vacuum. The matter source to be ablated was a cylindrical piece of 3 mm in height and 2 cm in diameter cut ~~ted~~ from an ingot prepared by melting the pure constitutive elements by induction under helium atmosphere in a copper mould crucible using r-f energy. TiNiZr-0 (respectively TiNiZr-1) were synthesized by depositing on a substrate whose temperature was kept at 538 K (respectively 298 K). The films were not further annealed. Films of around 0.6 μm thickness were obtained. More details concerning the deposition conditions can be found in [20].

Al-Cu-Fe samples were prepared by reactive cathodic magnetron sputtering. Two Al-Cu-Fe thick films of composition $Al_{62.5}Cu_{5.5}Fe_{12.5}$ were prepared. Al-Cu-Fe-1 was prepared on a glass substrate whose temperature was kept at 298 K. The disk target was made of three sectors, each sector being made of each of the elements (Al, Cu, and Fe). Background vacuum was 0.3 Pa, a current of 0.6 A was maintained on the sputtered target and the power used was 250 W. A thick film of around 6 μm was obtained. Al-Cu-Fe-0 is the Al-Cu-Fe-1 sample that has undergone ex-situ a thermal treatment of 2 h at 673 K. As thicknesses of the films are huge regarding to the usual coherency lengths of the quasicrystalline structures, they can be considered as bulk material and can thus be compared to the sintered ones.

## III. CHEMICAL, STRUCTURAL AND OPTICAL CHARACTERIZATION OF SAMPLES

Atomic compositions of the samples were checked by Electron Probe Microanalysis.

The structure of the samples was determined by X-Ray Diffraction ($CoK_\alpha$ radiation, $\lambda$ = 0.178897 nm). In case of a texture (typically for TiNiZr thick films[21]), a four-circles diffractometer was used. 22 scans were so recorded at different $\chi$ angles and added in order to cumulate and visualize all the possible Bragg reflections on a same diagram. This acquisition was done with $\chi$ ranging from 0 to 55° assuring a $\varphi$ rotation of the sample at 600 rotations per minute. Such an acquisition allows to visualize all the possible Bragg diffraction peaks in the range $30° \leq 2\theta \leq 96°$ (where $\theta$ is the Bragg angle). Quasicrystalline state was systematically checked by TEM on a PHILIPS CM200 microscope. Morphologies were studied by TEM (PHILIPS CM200) or by SEM on a PHILIPS XL-FEG microscope.

The bulk sintered samples were mirror polished before their reflectivity R could be measured (mechanical standard polishing using SiC paper grids, followed by diamond paste polishing was performed). The Ti-Ni-Zr and Al-Cu-Fe samples' reflectivity was measured on the as synthesized films. The reflectivity spectra were measured using a Perkin-Elmer spectrometer in the wide frequency range 4100 - 50 000 $cm^{-1}$ at nearly normal incidence. All samples, due to their preparation method, are very dense. Their porosity (far less than 10 %) makes them adequate samples to be optically characterized, notably in the UV domain.

## IV. EXPERIMENTAL RESULTS

### A. Optics

The complete set of optical reflectivity R data obtained with the Al-Cu-Fe-B and Al-Pd-Mn sintered samples and with the Ti-Ni-Zr and Al-Cu-Fe thick film samples is shown on the curves in Fig. 1. The authors have deliberately decided to present them versus the wave numbers instead of versus the wavelengths as this presentation brings out the behaviour in the UV domain. The optical data of the Al-Cr-Fe and Al-Cr-Fe-Mo approximants are given in Fig. 2. Whatever the samples, all reflectivity curves globally follow a plateau around 60 % in the infra-red and visible domain until around 25 000 $cm^{-1}$. Above this value, in the UV domain, reflectivity behaviour is not the same for all samples. The differences between samples allow a classification ~~of samples~~ in four groups. The first type of behaviour is a decreasing curve adopting a smoothed steps shape (Group A = AlCuFeB-0, AlPdMn-0). The second type of behaviour is a decreasing curve with some and very weak modulations (Group B = AlCuFeB-1, AlCuFeB-2, AlPdMn-1, AlCuFe-0, TiNiZr-0, TiNiZr-1). A third group is constituted by the alloys whose reflectivity curve roughly keeps the 60 % value over all the measured domain (Group C = Approx-0, Approx-1 and Approx-2). The fourth group is characterized by an increase of R over the UV domain (Group D = AlCuFe-1).

B. Morphological and structural characterization of samples

X-ray and TEM analyses showed samples are all crystallized (quasicrystalline or approximant) except AlCuFe-1 which is amorphous. Table 1 (respectively 2) compiles the structural and morphological details of the sintered (respectively thick film) samples.

*1. Sintered samples*

All sintered samples (Al-Cu-Fe-B, Al-Pd-Mn, Al-Cr-Fe and Al-Cr-Fe-Mo ones) exhibit granular morphologies whose grain size measured on SEM pictures is of the µm order. Fig. 3(a) shows the differences between the three Al-Cu-Fe-B samples prepared. The diagram of AlCuFeB-0 sample exhibits the characteristic peaks of the ~~pure~~ icosahedral phase as it was indexed, the peaks are well defined and well separated. The diagram of sample AlCuFeB-1 (very similar to the former one) attests the sample is mostly composed of the quasicrystalline phase with a residual CsCl-phase as the presence of the 51° peak shows it (as shown by the arrow on the figure). Such a pollution is due to the post-treatment temperature at 373 K which is not adequate to completely eliminate this phase. The diagram of sample AlCuFeB-2 is constituted of much less resolute and larger peaks than the diagram of AlCuFeB-0 (namely in the 36 - 47° range) testifying to ~~attesting~~ a poorer structural perfection than the one of AlCuFeB-0. Moreover, the diffuse scattering of the AlCuFeB-2 diagram localized around $2\theta = 52°$ is the result of structural disorder in the packing of atomic clusters in the icosahedral structure.

Identical qualitative differences can be spotted between the AlPdMn-0 and AlPdMn-1 samples (Fig. 3(b)). AlPdMn-0 is the sample owning the best icosahedral order as attested : on the one hand by a close to zero diffuse scattering observed at the bottom-foot of the peaks of the icosahedral phase (as indexed on the figure) and on the other hand by the thin width of peaks very close to instrument resolution only. AlPdMn-1 is an icosahedral sample with local structural disorder as confirmed by the two maxima of diffuse scattering ~~respectively~~ located at $2\theta = 34°$ and 51°.

Fig. 4 shows the X-ray diffraction diagrams of the approximant samples : Approx-0, Approx-1 and Approx-2 samples. They attest of a good crystallization of these samples and show that Approx-0 is made of a $\gamma$ brass phase, Approx-1 (Al-Cr-Fe) and Approx-2 (Al-Cr-Fe-Mo) are composed of an orthorhombic O1 phase. The $\gamma$ brass phase is an approximant of both the decagonal and the icosahedral phases and the $O_1$ phase is an approximant of the known decagonal phase. The $\gamma$ brass phase is a rhombohedral R3m structure with a = 0.7805 nm , and $\alpha = 109.7°$. The AlCrFe-$O_1$ phase is an orthorhombic Bmm2 phase with a = 3.25 nm, b = 1.22 nm, c =

2.36 nm. Comparison of diagrams of fig.4 (b) and (c) shows that the insertion of Mo in the system does not change the crystallographic parameters.

*2. Thick film samples*

The Ti-Ni-Zr samples X-ray diagrams are displayed in Fig. 5(a). Extensive details on the structure of these films were published in detail elsewhere[21]. TiNiZr-0 sample is made of quasicrystalline columns normal to their substrate. Their average width is 50 nm. The columns are textured along a fibre located around the five fold axis. TiNiZr-1 is homogeneous. It exhibits an icosahedral nanocrystalline order. Calculation thanks to Scherrer law applied on the full width at half maximum of deconvoluted peaks of the main signal of the diagrams [21] give 2 (±1) nm (micro-strain and lattice distortion were assumed to be nil). The grain size of these samples is of the order of the lattice parameters (few nm) of approximant phases which are the related crystalline phases of the quasicrystals. It implies the Ti-Ni-Zr samples can be considered either as icosahedral matter with grain size of few nm, or as approximant with a lattice parameter equal to the grain size.

Fig. 5(d) gives the X-ray diagrams of the Al-Cu-Fe samples. AlCuFe-1 sample is amorphous (attested by TEM experiment, in dark field images : no nanocrystal are observed). AlCuFe-0 is the AlCuFe-1 sample that has undergone a thermal treatment. This thermal treatment leads to a nanocrystallization of the film (attested by TEM dark field experiment). The average particle size measured is 3 (±1) nm. This could be obtained by direct measurement of the TEM images and was confirmed by calculation based on Scherrer law applied on deconvoluted peaks of the main XRD signal (micro-strain and lattice distortion were assumed to be nil).

## IV. DISCUSSION

The classification in four groups as well as the contradictory results of literature can actually be interpreted with the Model of Janot[23] and taking into considerations the crystallographic and morphological state of the samples. Indeed, the icosahedral structure (the one of Al-Pd-Mn, Al-Cu-Fe, Al-Cu-Fe-B and Ti-Ni-Zr) is a hierarchical arrangement of clusters where every hierarchical level is obtained by inflation of the one before and where each atomic cluster can be considered as a potential well. The model of Janot says 1) the resonance frequencies in the optical conductivity (optical conductivity and reflectivity data are mathematically linked by a Kramers Krönig transformation [15]) are the result of hierarchical variable-range electronic hopping mechanism between potential, or localization wells[14,23] 2) the electronic hopping distances Λ are indeed linked to the resonance frequencies $w_0$ by $\Lambda=\sqrt{\frac{2h}{mw_0}}$ , h being the Planck constant, m the mass of the free electron. In the quasicrystalline structure the most representative hopping distance is the one in between clusters from centre to centre (adjacent but not interpenetrating). Theses distances are 12 and 14 Å for the Mackay and the Bergmann clusters, respectively. The Mackay type cluster is found in the Al-Pd-Mn, Al-Cu-Fe and Al-Cu-Fe-B systems, the Bergmann type one in Ti-Ni-Zr. It results in an optical conductivity broad resonance centred around 15000 $cm^{-1}$. The UV domain starts around 25000 $cm^{-1}$. It implies that the effective hopping distances related to that domain are smaller : they correspond to electrons tunnelling from sites like from cluster centres to an atom being in flipping position [15] that is to an atom indifferently belonging to one of the two interpenetrated clusters. This study shows that the strong differences noted in reflectivity from one sample to another are located in this domain : they can be characterized and differentiated. They can also be enhanced by calculating the first-order derivatives of the optical curves : Fig. 6 shows the derivative optical curves of the quasicrystalline samples of the paper.

The canonical behaviour of a quasicrystalline alloy can be observed on the reflectivity curves of the samples of group A (Fig. 1) whose structure can be described as excellent (AlCuFeB-0, AlPdMn-0) : the observed drop of reflectivity goes off in several steps. The derivative curves (Fig.6(a) and (b)) indeed present two or three strong and net variations. Middle of each reflectivity step as indicated by arrows on the figure precisely corresponds to a maximal resonance of the conductivity[15]. The two steps centred at 41000 $cm^{-1}$ and around 52000 $cm^{-1}$ are associated to electronic jumps of 11 Å and 9.5 Å, respectively. They correspond to jumps from a cluster centre to an atom being in flipping position. This could explain why these steps could not be readable in some literature works although some irregularities on the log-log plot could be guessed. The characterized quasicrystalline alloys were probably not of sufficient structural quality.

When the order of the structure is perturbed as in the case of samples exhibiting X-ray diffuse scattering (sub group of group B : AlCuFeB-1, AlCuFeB-2, AlPdMn-1), the hierarchical geometrical arrangement of atomic clusters is modified. The deterioration of the structural perfection affects the densities and frequencies of electronic jumps, it decreases the conductivity resonance peaks and results in the disappearance of the stair behaviour of the reflectivity. The larger the diffuse scattering is, the less the peaks are defined and the more the distribution of structural distances spreads out. It results in a weaker decrease of reflectivity in the UV domain, and in the smoothing or disappearance of the steps (Fig. 1 (a) and (b)). This can be seen even more clearly on three derivative curves that are actually flat (Fig.6 (a) and (b)).

The second sub-group of the group B is made of samples whose grains get as small as a few nanometers : AlCuFe-0, size of grains = 3 nm , TiNiZr-0 size of grains = 50 nm, and TiNiZr-1 size of grains = 2 nm; the grains are thus around 100 times smaller than for the sintered samples (few μm). The first consequence is that the distribution statistics of the in between sites distances is altered. The second point is that the ratio of surface atoms over volume atoms is strongly modified, many clusters are broken, and do not act as potential wells, it drastically reduces the statistics of some electronic jumps. These two points result in a smoothing of the maxima in conductivity and of course of the steps in reflectivity and by the disappearance of the strong and abrupt variations in the derivative curves (Fig.6(c) and (d)). The optical reflectivity curves then decrease over the UV domain with no clear step behaviour (Fig. 1 (c) and d)), but with nearly non noticeable undulations reminding the step behaviour. This phenomenon is indeed progressive according to the decrease of the size of the grains. It can be seen on the curves of the samples of the second sub-group of the group B : TiNiZr-0 size of grains = 50 nm, TiNiZr-1 size of grains = 2 nm and AlCuFe-0, size of grains = 3 nm). The sample TiNiZr-0 is made of columnar grains perpendicular to the surface : the width of the columns (50 nm) is well the significant distance of the size of the grains as the optical waves will not be affected by the height of the columns (short penetration perpendicular to the surface). The derivative curves also confirm the evolution is progressive according to the size of the grains (Fig.6 (c)).

For the approximant samples which are crystalline (samples of group C : Approx-0, Approx-1 and Approx-2), the clusters do exist locally inside the cells, but the hierarchy (clusters of clusters) is non existant. The potential wells disappear, tunnelling occurs then from atoms to atoms. The hopping distances distribution is in this case a distribution more continuous than previously. Optical conductivity will show up no resonance peak, and the optical reflectivity curves will exhibit no step (Fig. 2). There is no need here to look at the derivatives of the curves. The plateau goes on over the UV domain. Although, local order of approximant material is very close to the one of quasicrystals, the loss of quasicrystallinity is stronger than for the quasicrystalline samples where the ~~Granularity~~ granularity is around a nanometer. The content of the residual presence of the β phase (β peak at 51°) which is an approximant of the AlCuFeB icosahedral phase in the AlCuFeB-1 sample is too small to have an important impact on the reflectivity curve which has already been altered by a smoothing due to local structural disorder.

For samples whose structure is amorphous (Group D = AlCuFe-1), no cluster exists. The material is a dense random packing of hard spheres (atoms). The sample does not possess any hierarchy beyond local atomic distances : atoms are ordered over lengths inferior to two or three atomic distances. This leads to a continuous distribution of possible electronic jumps and to no frequencies resonance. It also leads to a progressive increase of the frequencies corresponding to the smallest distances. The plateau goes over the UV domain, presents no step and is continuously increasing.

## VI. CONCLUSION

A study was performed to get the canonical icosahedral quasicrystalline, approximant and amorphous phases response in optical reflectivity over the UV domain. It is remarkable to realize that a net and easy to read differentiation of optical behaviour occurs according to very small structural “shifts” from a “good” quasicrystallinity. Indeed, loss of order such as the one noticeable by the discrete presence of diffuse diffusion background at the base of X-ray diffraction peaks or too small coherency lengths (due to the size of grains, or to the approximant structure) are sufficient to dramatically affect the UV optical reflectivity. Amorphous phases lead to an increase of the signal over this domain. The reflectivity measurement, and all the more its derivative curve, are thus very sensitive. The measurements can be performed quickly and cheaply : it could thus be used by manufacturers of quasicrystalline matter to test their samples quality. It can also help to make the difference between amorphous and quasicrystalline samples with small grains (nanometer range) as it is not possible to do so by looking only at X-ray diffraction diagrams. This is all the more important since one might expect similar sensitivities of  other properties of quasicrystals. Indeed optical properties, like many other properties of the quasicrystals take their foundation on the peculiar waves interactions inside the structure.

This work is a contribution to understand the relationship structure-properties of nanoscaled materials.

The study stresses on why, as already mentioned in the introduction, quality of samples is of high relevance to characterize canonical properties of materials. This is for quasicrystals here, apparently even more true.

## ACKNOWLEDGEMENTS

We thank G. Jeandel (LEMTA-CNRS-Nancy) for lending the optical characterization equipement, V. Demange and H. L Tran for preparing some samples and J. B. Ledeuil (LSG2M-CNRS-Nancy) for the EPMA analysis. The authors would like to thank the Région Lorraine and Saint-Gobain Recherche, as the French Ministry of Research and New Technologies for subsidizing the “ERT Quasicristaux Industriels”.

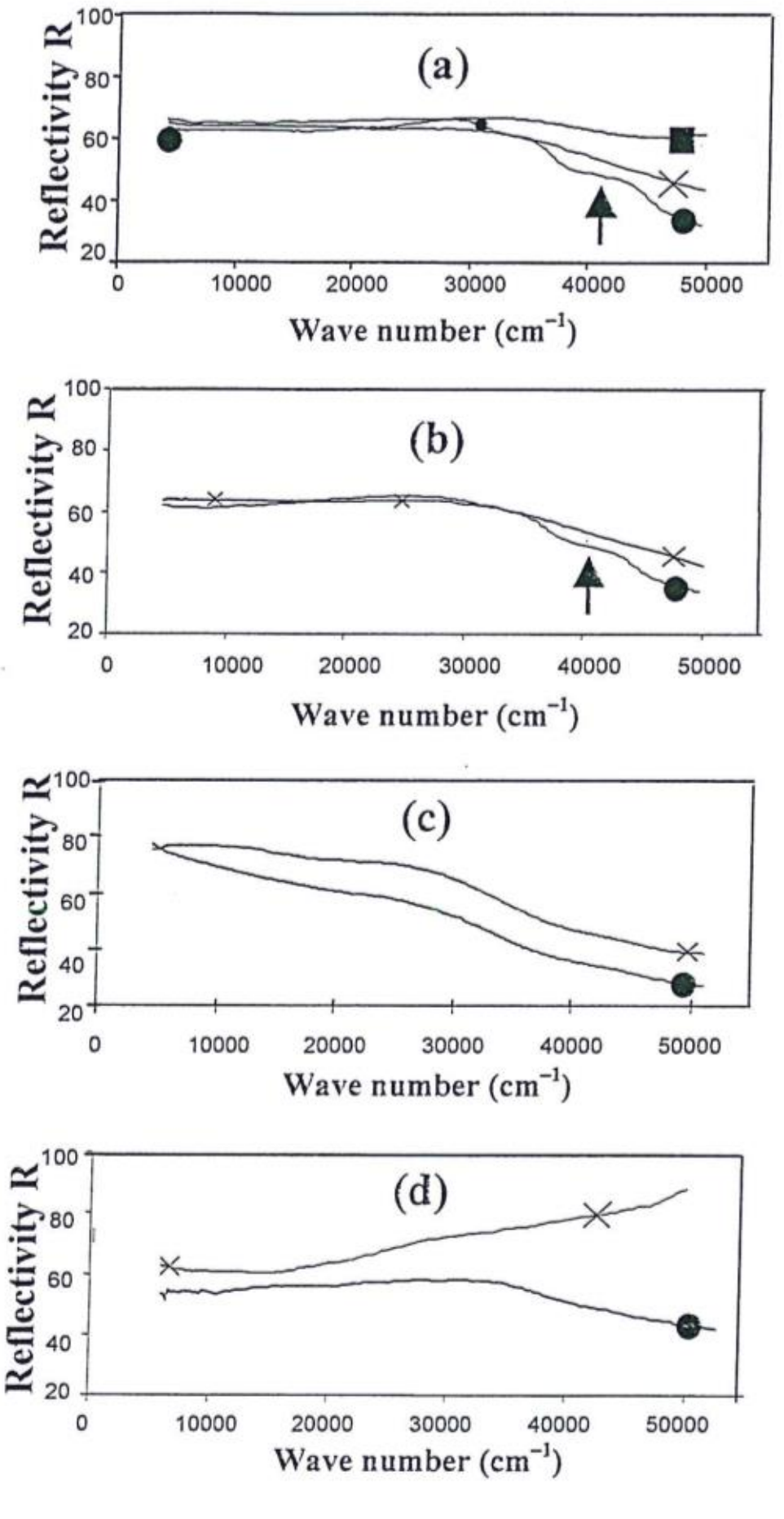

FIG. 1 Optical reflectivity versus wave number in the near infra red, visible and ultra-violet domain of quasicrystalline sintered samples : samples belong to different systems and exhibit different structural qualities. a/ Al-Cu-Fe-B system, b/ Al-Pd-Mn system c/ Ti-Ni-Zr system and d/ Al-Cu-Fe system. ●,x and ■ represent the X-0, X-1 and X-2sample, respectively, X standing successively for AlCuFeB, AlPdMn, TiNiZr and AlCuFe.

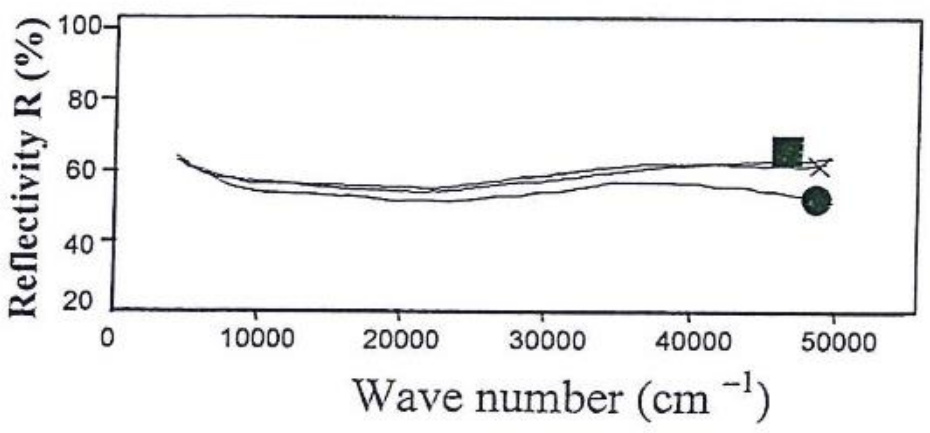

Figure 2

FIG. 2 Optical reflectivity versus wave number in the near infra red, visible and ultra-violet domain of approximant sintered samples. a/ Approx-0 b/ Approx-1 and c/ Approx-2. ●,x and ■ represent the Approx-0, Approx-1 and Approx-2 samples, respectively.

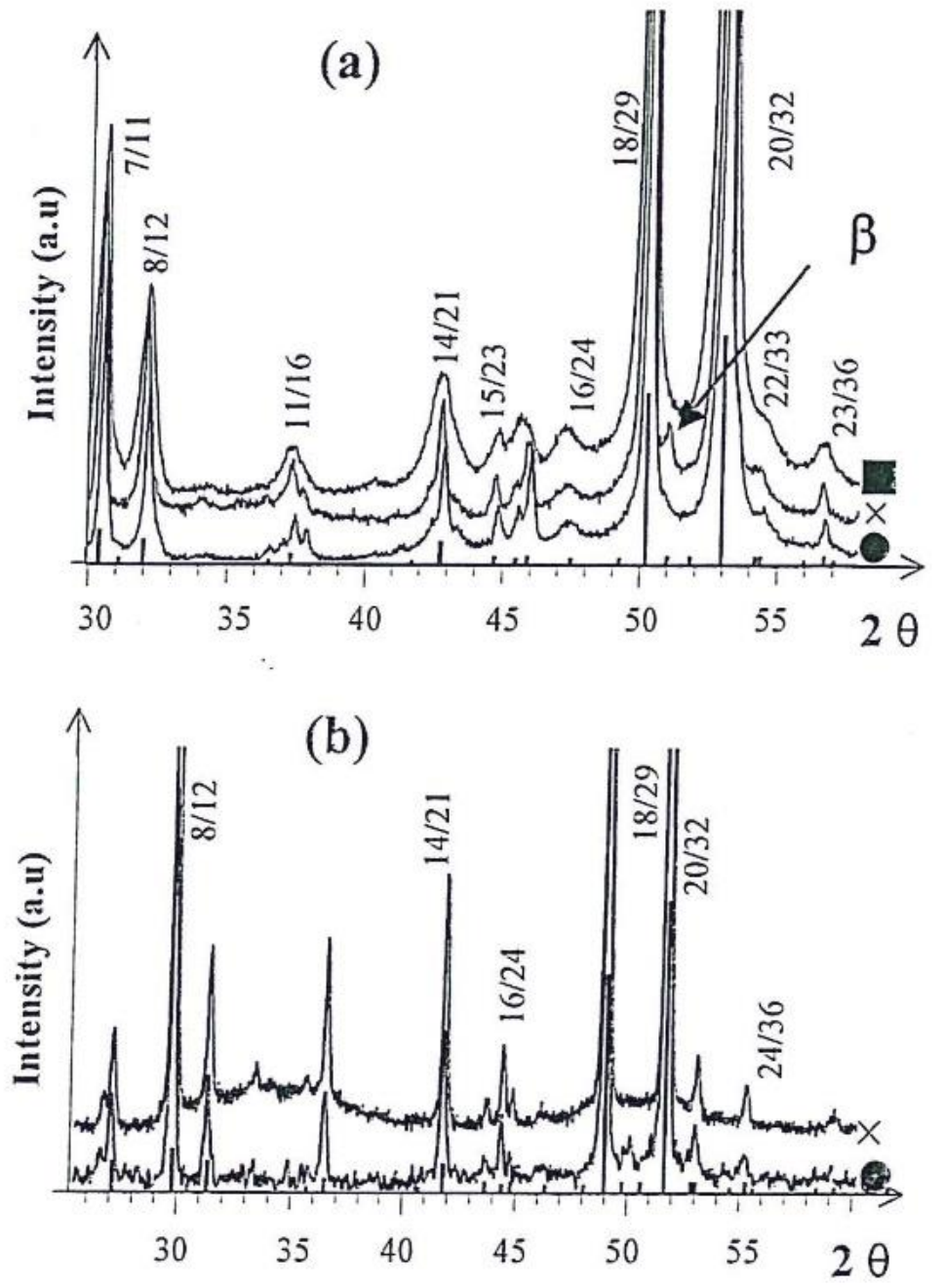

FIG. 3 X-ray diffraction diagrams ($\lambda_{CoK\alpha}$) of a/ Al-Cu-Fe-B samples and b/ Al-Pd-Mn samples. Diagrams have been zoomed on adequate 2θ ranges to allow the comparison between the diagrams since details are small. ●, x, and ■ represent the X-0, X-1 and X-2sample, respectively, X standing successively for AlCuFeB and AlPdMn.

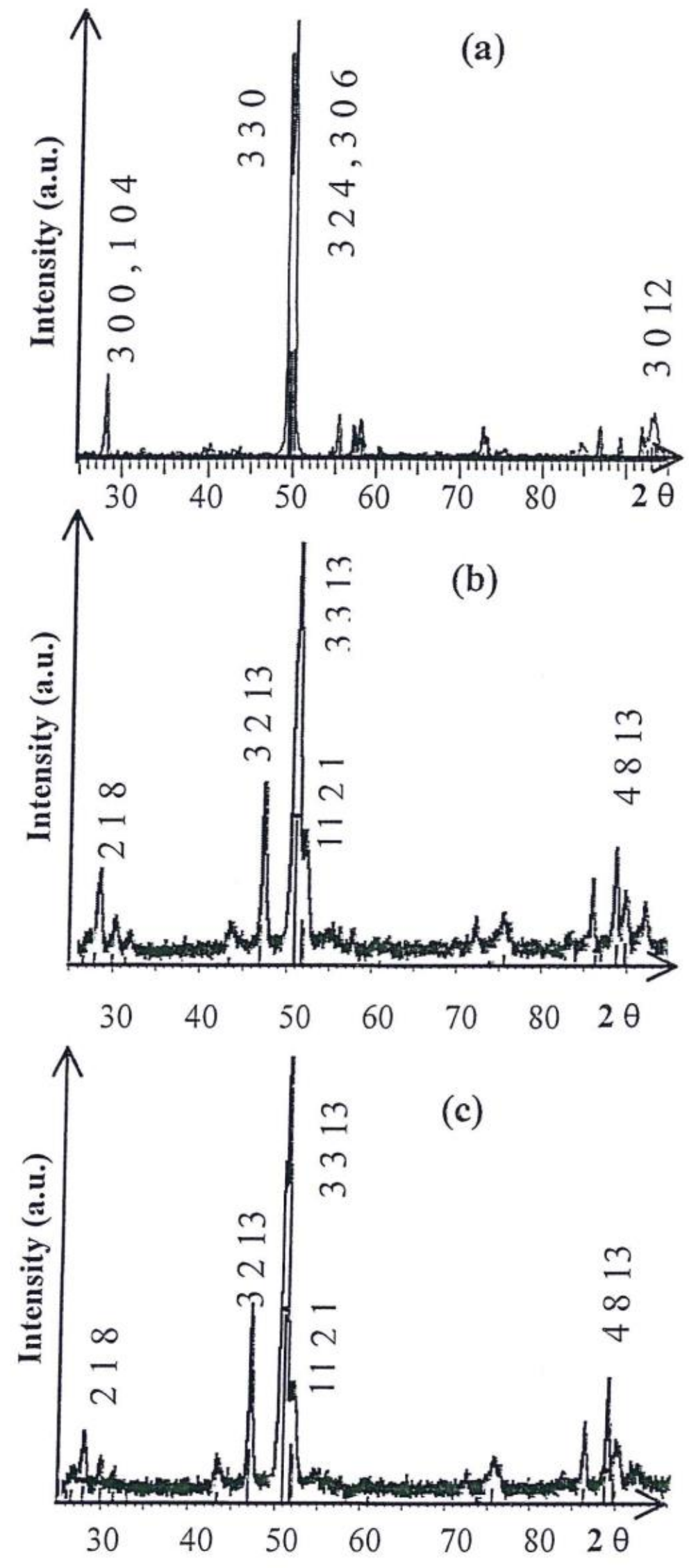

Fig. 4

FIG. 4 X-ray diffraction diagrams ($\lambda_{CoK\alpha}$) of a/ Approx-0, b/ Approx-1 and c/ Approx-2.

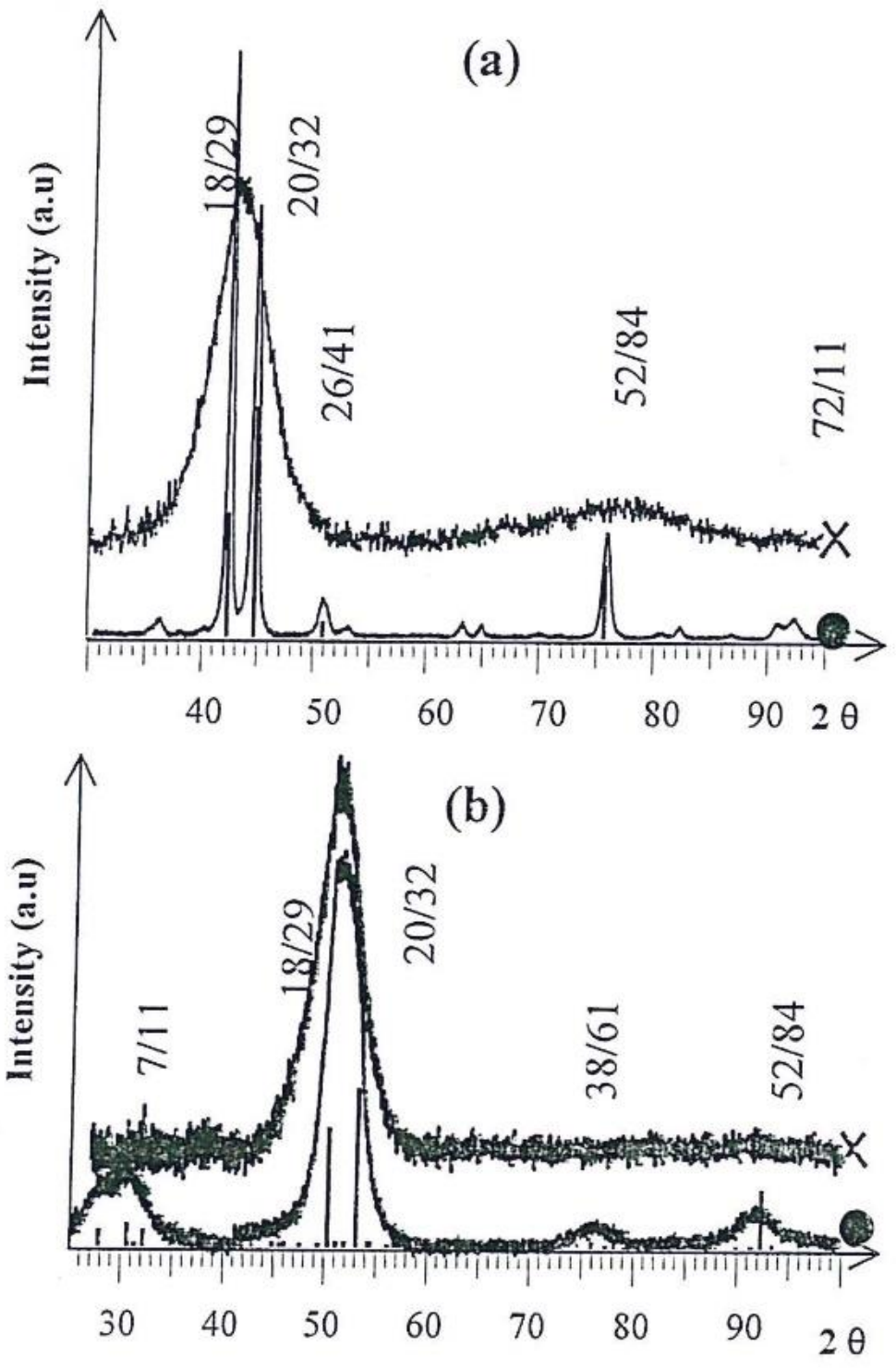

Figure 5

FIG. 5 X-ray diffraction diagrams ($\lambda_{CoK\alpha}$) of a/ Ti-Ni-Zr samples and b/ Al-Cu-Fe samples. ● and x represent the X-0 and X-1sample, respectively, X standing successively for TiNiZr and AlCuFe.

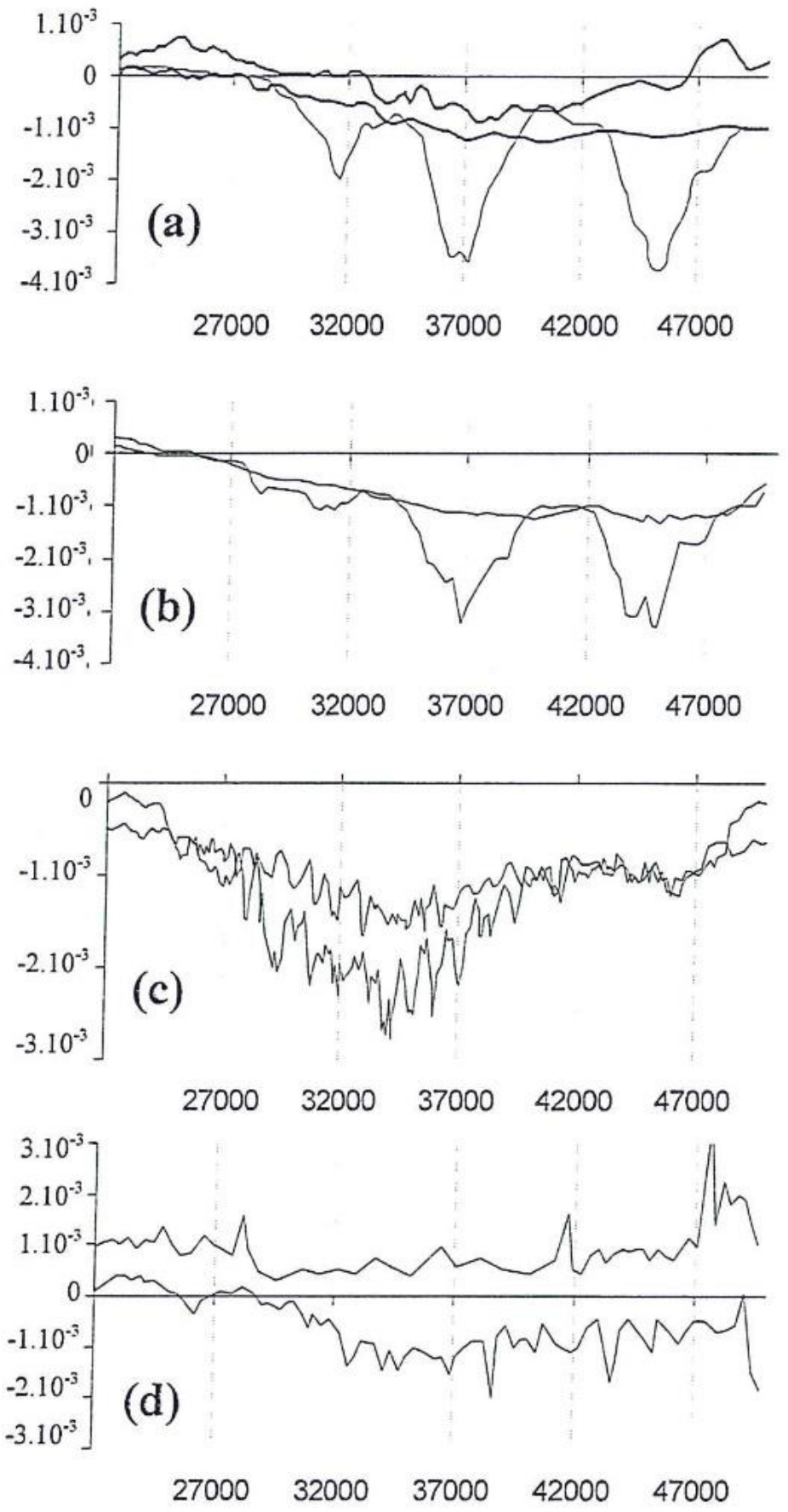

Figure 6

FIG. 6 First-order derivatives curves of reflectivity curves presented in fig. 1. a/ Al-Cu-Fe-B system, b/ Al-Pd-Mn system c/ Ti-Ni-Zr system and d/ Al-Cu-Fe system. ●,x and ■ represent the X-0, X-1 and X-2sample, respectively, X standing successively for AlCuFeB, AlPdMn, TiNiZr and AlCuFe.

TABLES

TABLE I. Synthesis experimental, structural and morphological details of sintered bulk samples. "diffuse diffusion" means of the structure is slightly and locally disordered attested by diffuse diffusion at the base of the peaks of the X-ray diffraction diagrams (Fig. 3).

| Sample<br>Composition<br>(atomic %) | Base material<br>Granularity of<br>powder (μm) | Sintering<br>treatment<br>Temperature (K)<br>Stress (MPa) | Post-treatment<br>Temperature (K)<br>Time (h) | Structure<br>Morphology<br>Size of grains |
|---|---|---|---|---|
| **AlCuFeB-0**<br>**$Al_{58.4}Cu_{25.1}Fe_{12.5}B_4$** | Atomized powder<br>25 | 830<br>40 | 780<br>3 | Icosahedral<br>Polycrystalline<br>Few μm |
| **AlCuFeB-1**<br>**$Al_{58.4}Cu_{25.1}Fe_{12.5}B_4$** | Powder from crushed ingots<br>25 - 65 | 780<br>15 | 100<br>3 | Icosahedral + diffuse diffusion + residual presence of CsCl<br>Polycrystalline<br>Few μm |
| **AlCuFeB-2**<br>**$Al_{58.4}Cu_{25.1}Fe_{12.5}B_4$** | Powder from crushed ingots<br>25 - 65 | 1053<br>15 | 903<br>3 | Icosahedral + diffuse diffusion<br>Polycrystalline<br>Few μm |
| **AlPdMn-0**<br>**$Al_{69.8}Pd_{21.6}Mn_{8.6}$** | Powder from crushed ingots<br>25 - 65 | 1103<br>15 | 873<br>3 | Icosahedral<br>Polycrystalline<br>Few μm |
| **AlPdMn-1**<br>**$Al_{69.8}Pd_{21.6}Mn_{8.6}$** | Powder from crushed ingots<br>25 - 65 | 1103<br>15 | 373<br>3 | Icosahedral + diffuse diffusion<br>Polycrystalline<br>Few μm |
| **Approx-0**<br>**$Al_9(Cr,Fe)_4$** | Powder from crushed ingots<br>25 - 65 | 1333<br>15 | 373<br>2 | γ brass phase<br>Polycrystalline<br>Few μm |
| **Approx-1**<br>**$Al_{71.6}Cr_{22.4}Fe_6$** | Powder from crushed ingots<br>25 - 65 | 1253<br>15 | 373<br>2 | $O_1$ phase<br>Polycrystalline<br>Few μm |
| **Approx-2**<br>**$Al_{71.6}Cr_{20.8}Fe_{5.9}Mo_{1.6}$** | Powder from crushed ingots<br>25 - 65 | 1253<br>15 | 373<br>2 | $O_1$ phase<br>Polycrystalline<br>Few μm |

TABLE II. Synthesis experimental structural and morphological details of thick films samples.

| Sample<br>Composition | Preparation technique<br>Thickness (μm) | Substrate temperature (K) | Post-treatment<br>Temperature (K)<br>Time (h) | Structure<br>Morphology<br>Size of grains |
|---|---|---|---|---|
| **TiNiZr-0**<br>**$Ti_{41.5}Ni_{17}Zr_{41.5}$** | PLD<br>0.6 | 548 | None | Icosahedral<br>Textured<br>Columnar,<br>50 nm (width) |
| **TiNiZr-1**<br>**$Ti_{41.5}Ni_{17}Zr_{41.5}$** | PLD<br>0.6 | 298 | None | Icosahedral<br>nanocrystalline<br>2 nm |
| **AlCuFe-0**<br>**$Al_{62.5}Cu_{5.5}Fe_{12.5}$** | Reactive magnetron<br>Sputtering<br>6 | 298 | 673<br>2 | Icosahedral<br>nanocrystalline<br>3 nm |
| **AlCuFe-1**<br>**$Al_{62.5}Cu_{5.5}Fe_{12.5}$** | Reactive magnetron<br>Sputtering<br>6 | 298 | None | Amorphous<br>-<br>- |